\documentclass[12pt]{iopart}
\usepackage{graphicx}
\usepackage{iopams}

\begin{document}

\title[]{Percolation induced effects in 2D coined quantum walks: analytic asymptotic solutions}

\author{B. Koll\'ar$^{1,2}$, J. Novotn\'y$^3$,  T. Kiss$^1$ and I. Jex$^3$}
\address{$^1$ Wigner RCP, SZFKI, Konkoly-Thege M. u. 29-33, H-1121 Budapest, Hungary}
\address{$^2$ Institute of Physics, University of P\'ecs, Ifj\'us\'ag \'utja 6, H-7624 P\'ecs, Hungary}
\address{$^3$ Department of Physics, Faculty of Nuclear Sciences and Physical Engineering, Czech Technical University in Prague, B\v rehov\'a 7, 115 19 Praha 1 - Star\'e M\v{e}sto, Czech Republic}
\ead{\mailto{kollar.balint@wigner.mta.hu}}

\begin{abstract}
Quantum walks on graphs can model physical processes and serve as
efficient tools in quantum information theory. Once we admit random
variations in the connectivity of the underlying graph, we arrive at
the problem of percolation, where the long-time behaviour appears
untreatable with direct numerical methods. We develop novel analytic
methods based on the theory of random unitary operations which help us to
determine explicitly the asymptotic dynamics of quantum walks on 2D
finite integer lattices with percolation. Based on this theory we
find new unexpected features of percolated walks like asymptotic
position inhomogeneity or special directional symmetry breaking.
\end{abstract}


\maketitle

 \section{Introduction}
The dynamics of particles is one of the central problems of physics. The rules formulated for the motion of particles reflect our knowledge at an elementary level.
Surprisingly, on a larger scale we can recover certain properties of complex physical systems with models based on simple rules.

Recently, a novel elementary model of motion has been introduced for
quantum particles: quantum walks \cite{Aharonov1993,Kempe2003,Elias2012} have
been defined in analogy to classical random walks. Classical walks
reflect the fact that the motion of particles is given by sequences
of tiny collisions with inherently random character. Quantum walks (QWs)
are describing the motion of a quantum particle on a graph
 and capture its wave character \cite{Konno2008}.
The particle wave function is repeatedly split, the partial waves
mutually interfere forming a characteristic interference pattern.
This pattern is radically different from the classical counterpart
leading to rich behaviour from exponential speedup to localization.
In turn, these may lead to interesting physical effects in models of 
coherent transport. 
Much like their classical counterpart, QWs are suitable tools for computation, thus are perfectly fitting for
designing quantum algorithms \cite{Shenvi2003,Childs2003,Ambainis2007,Qiang2012,Paparo2012,Sanchez2012}. Moreover, QWs proven to be universal computational primitives \cite{Childs2009,Lovett2010}.
Due to their
potential, QWs attracted considerable experimental interest. A
number of experiments demonstrating QWs have been reported
with trapped atoms \cite{Karski2009,Alberti2013}, trapped ions \cite{Schmitz2009,Zahringer2010}, photons in optical waveguides \cite{OBrien2010,Silberberg2012,Mataloni2013,Thompson2013}, in an optical realization \cite{Broome2010} and in optical feedback loops \cite{Schreiber2010,Schreiber2011,Torres2012,Schreiber2012}.

Any realistic model has to respect the simple fact that the world
around us differs to a certain degree from the models we form to
describe it. These differences can be viewed as perturbations or
imperfections. While imperfections are usually expected to destroy
the ideal situation, on the other hand they can lead to completely
new effects which are not present in the original model. Quantum
walks are in this respect not an exception
\cite{Kendon2007,Romanelli2005,Torma2002,Shikano2010}. When focusing on the so
called discrete quantum walks we can study the ideal situation and
then analyze the effect of different perturbations. They can lead to
diffusion like behaviour representing the final state in a transition
from quantum ballistic motion to classical diffusion
\cite{Kendon2007}. When static phase randomness is present, the
quantum walk exhibits Anderson type localization. This effect was demonstrated experimentally
\cite{Schreiber2011,Mataloni2013}.

In the present paper we will focus on perturbations of the
underlying graph. Defects of the graph can be viewed as a classical
percolation problem combined with quantum dynamics
\cite{Romanelli2005,Kendon2010,ChandrashekarPrePrint}. Since
continuous time quantum walks serve as a model for electron
transport \cite{Mulken2011} these randomization of the underlying
graph occurs naturally as a model for disordered media
\cite{Mulken2010}. Even fairly small changes in the graph may lead
to radical alteration of the dynamics \cite{Xu2012}. Static
percolation problem (where the disorder does not change througout the time evolution) for the continuous time quantum walks is
sometimes referred to as: statistical networks
\cite{Mulken2007,Anishchenko2012}. Dynamical percolation, i. e.
changing the graph during the time evolution repeatedly, has
recently been considered for the continuous time case
\cite{Darazs2013}. Discrete quantum walks are usually defined on a
regular lattice and the motion of the particle is defined by the
evolution operator which is the product of the coin and the step
operator. The defects of the graph determine the form of the evolution operator \cite{Kendon2010}.

In a previous paper \cite{Kollar2012} we introduced a general
formalism describing the evolution of quantum walks on graphs with dynamically occurring
defects -- percolation graphs
\cite{Percolation,Percolation2,DynamicalPercolation}.
We applied the general method of random unitary operations
in order to determine the possible asymptotic states.
This approach allowed us to identify the so-called
attractor space and to specify the long time behaviour of the system.
Beside the completely mixed state we found other stationary states as well
as asymptotically oscillating solutions.
The application of these results to higher dimensional graphs is not a trivial one:
 while the general formalism is in principle independent from the dimension
of the underlying graph, in practice carrying out the prescribed calculations
directly is very involving, seriously limiting its use for more complicated cases.

In this paper we improve significantly our method which, in certain cases, leads to a
considerably more efficient way to construct the attractor space and
the asymptotic dynamics and is particularly suited for dynamically percolated
quantum walks on integer lattices. We point out that static percolation is not the
subject of our treatment. We demonstrate that under rather
general conditions one can construct the attractor space using pure
quantum states. In this way we have a much better physical insight
into the asymptotic behaviour through the analysis of the properties
of the pure states constituents.

Coined quantum walks in two dimensions exhibit a number of interesting
effects \cite{Mackay2002} which are not present in the one-dimensional systems. A prominent
example is localization or trapping at the origin \cite{Tregenna2003,Inui2004,Stefanak2008a,Stefanak2008b}
for an initially spatially localized walker. Here we report on
a number of novel features for two-dimensional quantum walks on percolation lattices.
We illustrate the rich spectrum of properties of two-dimensional
percolated walks for three special types of coins. The first one is the
Hadamard coin extended to two-dimensional walks. In striking contrast to the
one-dimensional case, here a nonuniform position distribution can
be obtained asymptotically. Another new effect is the breaking of the directional symmetry:
Should we rotate the initial state and the underlying graph by 90 degrees in a certain
unpercolated walk, the obtained position distribution will also be rotated. However, if
we add percolation, then the above directional symmetry will be broken.
Next, we show that for percolated walks driven by the Grover coin trapping
(localization) can be found.
 Our last example is the Fourier coin, where we prove that the asymptotic position
 distribution is always uniform independent of the initial state.

All the obtained results shed light on open system dynamics in
connection with quantum walks and are of theoretical interest
presenting novel mathematical methods allowing for unexpected
results. In addition, the present results are not of purely
theoretical value but form the basis for experimental
implementations. In view of recent developments in optical
realization of two-dimensional walks \cite{Schreiber2012} the
predicted effects could be experimentally observed.

The paper is organized as follows. First, we introduce our notation
and define the model of quantum walks on percolation graphs. Then,
we review the general method for obtaining the asymptotic states of
the model. In section \ref{purestatemethod}. we present a procedure using the
pure eigenstate ansatz. In the next section we work out the case
of quantum walks on 2D percolation graphs in detail and present
three explicit examples: the Hadamard walk, the Grover walk and the
Fourier walk. Finally, we draw some conclusions.

\section{Definitions}

Given a $d$-regular finite simple graph $G(V,E)$, the Hilbert space
$\mathcal{H}$ of a discrete time quantum walk is defined as the
tensor product of the position space $\mathcal{H}_P$ and the coin
space $\mathcal{H}_C$. Here, the position space is spanned by state
vectors corresponding to the vertices of the graph $G$, whereas the
coin space is spanned by state vectors corresponding to the
directions of possible nearest neighbour hops between adjacent
vertices, forming a $d$-dimensional space.

\begin{figure}[tb!]
\begin{center}
\includegraphics[width=0.85\textwidth]{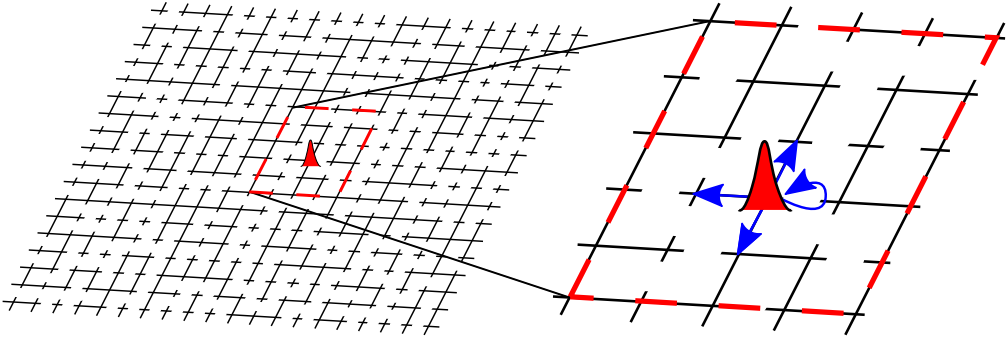}
\end{center}
\caption{
A 2D percolated grid with a quantum walker (red wave packet) and a magnified image of its immediate neighbourhood (right).
The link to the right is missing  ($ | R \rangle$ direction), and the walker cannot step in that direction.
To keep the QW
unitary on such imperfect graphs, we introduce a reflection
operator $R$ -- represented by the blue loop on the right plot. The steps to all other directions are represented by the blue straight arrows.
}
\label{percolation_figure}
\end{figure}

We add dynamical edge (bond) percolation to the model. Roughly speaking, the graph of the walk is randomly changing at each step of the walk.  Every edge of
the graph $G$ has an independent, but the same $p$ probability of being
perfect (present), and the complementary $1-p$ of being broken
(missing). Such a percolation graph is shown in figure \ref{percolation_figure}.
Here we note, that although the presented methods are valid more generally, in this paper we will treat only 2D integer lattices with lattice size greater than $2$ in each directions.
Preceding every discrete step of a QW
we randomly choose an edge configuration  $\mathcal{K} \subseteq E$,
thus making the percolation change in time (dynamical percolation).
This dynamical percolation serves as a source of decoherence, induced
by classical randomness --- lack of control over the system.

One can ask immediately the following question: how do we define the
actual time evolution of a QWs on an imperfect (percolation) graph? Our
approach is to keep the time evolution unitary. We introduce a
reflection operator $R$ to alter the definition of step (shift)
operator $S$ of the walk. The role of the reflection is simple:
whenever the walker faces a broken edge, instead of stepping
through, it undergoes a reflection in its internal coin degree of
freedom (see figure \ref{percolation_figure}.). This reflection is
carried out by the action of the reflection operator $R$. Employing
all these criteria, we can finally define a $\mathcal{K}$ (edge
configuration) dependent unitary shift operator
\begin{equation}
\fl S_\mathcal{K}   =  \sum_{(\mathbf{m}, \mathbf{m}\oplus c)  \not{\in} K}   | \mathbf{m} \rangle_P\langle \mathbf{m} |_P \otimes | c \rangle_C\langle c|_C R + \sum_{(\mathbf{m}, \mathbf{m}\oplus c)  \in K}  | \mathbf{m} \oplus c \rangle_P\langle \mathbf{m} |_P \otimes | c \rangle_C\langle c|_C\,,
\end{equation}
where $\mathbf{m} \oplus c$ denotes the nearest neighbour of graph vertex $\mathbf{m}$ in the direction labeled by coin state $c$.
Thus, the unitary time evolution of a single step on a given edge configuration $\mathcal{K}$ (imperfect graph) has the form of
\begin{equation}
 U_{\mathcal{K}} = S_\mathcal{K} \cdot (I_P \otimes C)\,,
\end{equation}
where $I_P$ is the identity operator acting on the position space,
and $C$ is the (special) unitary coin operator acting only on the
internal degree of freedom. The usage of $I_P$ reflects the fact,
that in this paper we treat the case where the system is
homogeneous, i. e. at every vertex the coin operator is the same.

From the building blocks we defined above, the full time evolution
can be constructed. The change of the underlying percolation graph after
each iteration is caused by classical randomness, thus it introduces
mixing during the time evolution. Hence, we use the density operator
formalism to describe the complete state $\hat{\rho}(t)$ of the walk
at discrete time $t$. The complete time evolution takes the form of
\begin{equation}
\hat{\rho}(t+1) = \sum_{\mathcal{K}} \pi_{\mathcal{K}} (p) U_{\mathcal{K}} \hat{\rho}(t) U_{\mathcal{K}}^{\dagger} \equiv \Phi( \hat{\rho}(t) )\,,
\end{equation}
where we introduced the linear superoperator $\Phi()$. With $\pi_{\mathcal{K}} (p)$ we denote the probability that the actual configuration $\mathcal{K}$ occurs.
By its construction, the superoperator $\Phi()$ belongs to the family of random unitary operations --- RUO maps.

\section{Asymptotics --- General method}
\label{generalmethod}

The long time behaviour rendered by such RUO maps are studied in detail in \cite{Novotny2009,Novotny2010,Novotny2012}.
Asymptotic states and their dynamics are governed by the so-called attractor space spanned by operators $X_{\lambda, i} \in \mathcal{B} (\mathcal{H})$, where all $X_{\lambda, i}$ satisfy the following set of equations
\begin{equation}
X_{\lambda, i} U_{\mathcal{K}} = \lambda  U_{\mathcal{K}} X_{\lambda, i}\,,
\label{attractor_space_matrices}
\end{equation}
where $|\lambda| = 1$ for all $\mathcal{K} \subseteq E$.
With $\mathcal{B} (\mathcal{H})$ we denote the space of linear
operators acting on the Hilbert-space $(\mathcal{H})$ of the system
in question. Note that $X_{\lambda,i}$-s are not necessarily valid
density matrices. The search for the attractors is usually quite
tedious. However, for certain cases a shortcut exists and the
attractor space can be constructed with the help of pure
eigenstates, which we discuss Section \ref{purestatemethod}.

The asymptotics can be readily determined with the following formula:
\begin{equation}
\hat{\rho}_{as} (t \gg 0) = \sum_{|\lambda|=1;\,i} \lambda^{t} X_{\lambda, i } \cdot \mathrm{Tr} \left( \hat{\rho}(t=0) X_{\lambda, i}^{\dagger} \right) \,,
\label{asymptotic_density_operator}
\end{equation}
where the phases of $\lambda$ are responsible for the appearance
of non-monotonous asymptotic dynamics, e. g. limit cycles. Moreover, we
require that $X_{\lambda,i}$-s form an orthonormal basis. In
general, the asymptotic behaviour of dynamics generated by RUO maps
is independent of the probabilities by which unitary operations are
applied, except for the extremal cases when $p=0$ or $p=1$
\cite{Novotny2010}.

For percolated QWs the very demanding work of finding the attractor
space matrices (\ref{attractor_space_matrices}) can be
simplified to a great extent with the use of the method proposed in
our previous paper \cite{Kollar2012}. Let us summarize the essential
main steps.

We rewrite conditions set by (\ref{attractor_space_matrices}) into the separated form
\begin{equation}
(I_P \otimes C) X_{\lambda, i } (I_P \otimes C^{\dagger}) \lambda^{*} =  S_{\mathcal{K}}^{\dagger} X_{\lambda, i } S_{\mathcal{K}}\,.
\label{separated_conditions}
\end{equation}
This form allows us to gain the solution in two steps. We have to find a subspace defined by shift operators corresponding to different configurations:
\begin{equation}
S_{\mathcal{K'}}^{\dagger} X_{\lambda, i } S_{\mathcal{K'}} =  S_{\mathcal{K}}^{\dagger} X_{\lambda, i } S_{\mathcal{K}}
\label{shift_conditions}
\end{equation}
for all $\mathcal{K'}, \mathcal{K} \subseteq E$.
We call the conditions imposed by shift operators as shift conditions, i. e. conditions connecting coin blocks (submatrices corresponding to the same position values) of attractor space matrices $X_{\lambda, i }$. For QWs on graphs with a high degree of symmetry these conditions are rather simple.

Moving forward, on the subspace determined by the shift conditions
only a single equation of type (\ref{separated_conditions})
should be fulfilled. The most convenient choice is the graph where
all the edges are missing. This results in the following local
condition
\begin{equation}
 (I_P \otimes RC)  X_{\lambda, i}  (I_P \otimes C^{\dagger} R^{\dagger}) = \lambda  X_{\lambda, i} \,.
 \label{coin_condition}
\end{equation}
Note, that the latter condition acts only on local coin subspaces,
and is homogeneous spatially. Thus, it can be decomposed into a set
of equations acting only on coin blocks of $ X_{\lambda, i}$
attractor matrices. The importance of this coin condition is
twofold: First, it determines the actual shape of the attractor
space matrices. Second, the $| \lambda| = 1$ eigenvalues are
determined by this equation. Therefore, one arrives at the
conclusion, that the form of the attractor space is highly dependent on
the coin operator $C$. The topology of the graph is reflected by the
shift conditions (\ref{shift_conditions}), while the coin condition
of (\ref{coin_condition}) is responsible for the appearance of
specific asymptotic behaviour, e. g. limit cycle.

The  $|\lambda|=1$ eigenvalues of (\ref{attractor_space_matrices}) are responsible for the type of
asymptotic dynamics which build up i.e. monotonous or limit cycle.
While a rich variety of asymptotic behaviours are available in the
full quantum state picture, the asymptotic position density operator
(and position distribution) of the QWs under consideration are
strictly stationary in time. This can be seen readily.

From equations (\ref{attractor_space_matrices}) and
(\ref{asymptotic_density_operator}) it is apparent that for all
asymptotic  $\hat{\rho}(t)_{as}$ quantum states a set of equations
\begin{equation}
\hat{\rho}(t+1)_{as} = U_{\mathcal{K}} \hat{\rho}(t)_{as} U_{\mathcal{K}}^{\dagger}
\end{equation}
for all $\mathcal{K} \subseteq E$
holds. Thus, the equation describing evolution on a graph with all
edges broken
\begin{equation}
\hat{\rho}(t+1)_{as}  =  U_{\{\}} \hat{\rho}(t)_{as} U_{\{\}}^{\dagger}  =   I_P \otimes RC_C \hat{\rho}(t)_{as} \left( I_P \otimes RC_C \right)^{\dagger}
\end{equation}
is also satisfied. Note, that this last expression is local: parts
of the density operator corresponding to different graph vertices
can not interfere. Thus, all asymptotic position density operators
satisfy the following equation:
\begin{eqnarray}
\fl \left( \hat{\rho}(t+1)_{as} \right)_P & \equiv & \mathrm{Tr}_C \left( \hat{\rho}(t+1)_{as} \right) \nonumber\\
& = &   \mathrm{Tr}_C \left\{\left( I_P \otimes RC_C\right) \hat{\rho}(t)_{as} \left( I_P \otimes RC_C \right)^{\dagger} \right\} \nonumber\\
& = &   \mathrm{Tr}_C  \left( \hat{\rho}(t)_{as} \right) \equiv \left( \hat{\rho}(t)_{as} \right)_P\,.
\end{eqnarray}
Therefore, all  asymptotic position density operators --- thus, the
asymptotic position distributions --- do not change with
additional interactions: the spatial distribution is stationary.

The general conditions (\ref{shift_conditions}) and (\ref{coin_condition}) considerably simplify the finding of asymptotics. However, especially for higher dimension this is still a formidable task. The problem can be significantly simplified by the following procedure.

\section{Pure eigenstate ansatz}
\label{purestatemethod}

Let us assume that we have the basis $\{ | \phi_{\alpha, i_{\alpha}} \rangle \}$ of common eigenstates satisfying
\begin{equation}
U_{\mathcal{K}} | \phi_{\alpha,  i_{\alpha}} \rangle = \alpha | \phi_{\alpha,  i_{\alpha}} \rangle
\end{equation}
for all $\mathcal{K}$-s,
where we use the index $i_\alpha$ to respect the possible degeneracy
of the eigenvalue $\alpha$. It is straightforward to see, that any
matrix from the span of $\{ | \phi_{\alpha,  i_{\alpha}} \rangle
\langle \phi_{\beta, i_{\beta}} | \}$ with a fixed eigenvalue
product $\alpha^{*} \beta = \lambda$, i. e.
\begin{equation}
Y_{\lambda, i} = \sum_{\alpha^{*} \beta = \lambda, i_{\alpha}, i_{\beta}} A^{\alpha, i_{\alpha}}_{\beta, i_{\beta}}  | \phi_{\alpha,  i_{\alpha}} \rangle \langle \phi_{\beta, i_{\beta}} |\,,
\label{pureattractors}
\end{equation}
is an attractor corresponding to the superoperator eigenvalue
$\lambda$ with matrix elements $A^{\alpha, i_{\alpha}}_{\beta, i_{\beta}}$ . Let us call these attractors p-attractors.

One can notice, that p-attractors in the form (\ref{pureattractors})
satisfy equations
\begin{equation}
Y_{\lambda, i} U_{\mathcal{K}} = \lambda  U_{\mathcal{K'}} Y_{\lambda, i}  
\label{pureattractor_conditions}
\end{equation}
where $|\lambda| = 1$ for all $\mathcal{K, K' \subseteq E}$.
This equation in comparison with the condition for general attractor
matrices (\ref{attractor_space_matrices}) is more restrictive.
Remarkably, the opposite implication is also true: Whenever a
nonzero matrix $X$ satisfies all equations
(\ref{pureattractor_conditions}), then $X$ always can be written in
the form (\ref{pureattractors}) with an $|\lambda| = 1$
superoperator eigenvalue.

The first statement is trivial and to prove the second statement we
first rewrite the set of equations (\ref{pureattractor_conditions})
into the vector form
\begin{equation}
U_{\mathcal{K}} \otimes U_{\mathcal{K'}}^{*} | y_{\lambda, i} \rangle = \lambda |y_{\lambda, i} \rangle\,,
\label{vectorized_pure_attractors}
\end{equation}
with $\langle p,q |y_{\lambda, i} \rangle \equiv \langle p
|Y_{\lambda, i}| q \rangle$ expanded in the natural (position $\otimes$ coin) basis.
Multiplying both sides by the probabilities of the unitaries
occurring during time evolution $\pi_{\mathcal{K}} (p),
\pi_{\mathcal{K'}} (p)$ and summing over all unitaries we transform
(\ref{vectorized_pure_attractors}) into an eigenvalue problem
\begin{equation}
B \otimes B^{*} |y_{\lambda,i} \rangle = \lambda |y_{\lambda,i} \rangle\,,
\label{pure_eigen_equation}
\end{equation}
with $B \equiv \sum_i \pi_{\mathcal{K}} (p) U_{\mathcal{K}}$ and $|\lambda|=1$.
The spectral radius of the map $B$ is bounded by one, the restriction of this map to Hilbert subspace corresponding to eigenvalues $|\alpha| =1 $ is diagonalizable and any eigenvector corresponding to an eigenvalue $|\alpha| =1 $ is a common eigenstate of unitaries $U_{\mathcal{K}}$ and vice versa. Hence, a general solution of (\ref{pure_eigen_equation}) is
\begin{equation}
|y_{\lambda, i} \rangle = \sum_{\alpha \beta^{*}=\lambda, i_{\alpha}, i_{\beta}} A^{\alpha, i_{\alpha}}_{\beta, i_{\beta}} |\phi_{\alpha, i_{\alpha}}\rangle \otimes \left( |\phi_{\beta, i_{\beta}} \rangle\right)^{*},
\end{equation}
which in matrix notation takes the form (\ref{pureattractors}).

We can conclude that an attractor $X_{\lambda, i}$ satisfying the set of equations $X_{\lambda, i}U_{\mathcal{K}}  = \lambda U_{\mathcal{K}}  X_{\lambda, i}$ can be constructed from common eigenstates, if and only if an attractor $X_{\lambda, i}$ satisfies the more restrictive set of equations $X_{\lambda, i} U_{\mathcal{K}}  = \lambda U_{\mathcal{K'}} X_{\lambda, i} $.
A direct consequence follows immediately: The trivial attractor proportional to the identity operator is clearly an attractor but not a p-attractor (except in the trivial case when a RUO map describes unitary evolution). Thus, the space of attractors always contains --- as a minimal subspace --- the span of p-attractors and identity. Surprisingly, in number of nontrivial cases this minimal subspace forms the whole attractor set and the asymptotic dynamics can be
analyzed readily.

Indeed, one can assume an orthogonal projection $\mathcal{P}$ onto
the subspace of common eigenstates of unitaries
$\{U_{\mathcal{K}}\}$. Let $\tilde{\mathcal{P}}$ be its orthogonal
complement satisfying $\mathcal{P} + \tilde{\mathcal{P}}=I$. In all cases, both projections are fixed points of the RUO
map $\Phi()$. Thus, the asymptotic dynamics after a sufficient
number of iterations  can be written as
\begin{equation}
\hat{\rho}(t \gg 1)  =  U_{\mathcal{K}}^t \mathcal{P}
\hat{\rho}(t=0) \mathcal{P}
\left(U_{\mathcal{K'}}^{\dagger}\right)^t + \tilde{\mathcal{P}}
\frac{\mathrm{Tr} \left\{ \hat{\rho}(t=0)
\tilde{\mathcal{P}}\right\}}{\mathrm{Tr} \tilde{\mathcal{P}}}
\label{asymptotic_unitary_evolution}
\end{equation}
for all unitaries corresponding to all pairs of indices
$\mathcal{K},\mathcal{K'}$. The asymptotic evolution in this case is
an incoherent (statistical) mixture of a purely unitary dynamics
inside the common eigenstates and a maximally mixed state living on
the orthogonal subspace.

With the just proposed procedure one can find elements of attractor space for a broad class of RUO time evolutions.
Computation of
p-attractors is much more easier usually than the computation of general
attractors, since finding common eigenvectors is an easier task than to find
general elements of the attractor space.
For arbitrary dimensional graphs the number of equations determining the general attractors scales quadratically with the size of the problem (dimension of the Hilbert space) while the proposed method scales linearly. In addition, the process of searching for attractors are more intuitive.
The common eigenstates also form a decoherence
free subspace, moreover they carry a physical meaning --- as elements of the Hilbert space --- while the general attractors not neccesarily have this property --- as they are elements of $\mathcal{B}(\mathcal{H})$.
In the following, we show that 2D percolated quantum walks can be successfully
treated with the just described pure state method.

\section{2D quantum walks}
\label{2dqws}

Let us employ our method now on the case of a 2D QW on a percolation lattice.
We will work out three specific examples of coin operators.
We consider 2D Cartesian (square) lattices (see figure
\ref{percolation_figure}.) with two different boundary conditions.
We investigate the cases of $M \times N$ tori (periodic
boundaries) and carpets (reflecting boundary conditions).
We note that the method presented in previous sections is
applicable also for other types of boundary conditions.

On percolated 2D systems the coin space corresponding to the
possible shift directions are spanned by the vectors $| L
\rangle_C,\,| D \rangle_C,\,| U \rangle_C,\,| R \rangle_C$.
The unitarity of the time evolution on an imperfect graph is ensured by the use of the
reflection operator. Throughout this paper, we use the
transposition matrix $\sigma_x \otimes \sigma_x$ as reflection
operator $R$, and we choose the coin matrix $C$ from the $SU(4)$ group.
Let us introduce a shorthand $|x, y, c\rangle \equiv
|\mathbf{m}=(x; y) \rangle_P \otimes |c\rangle_C$ for denoting state
vectors living on the composite Hilbert-space $\mathcal{H}$ of 2D
QWs.

We apply the general method reviewed in Section \ref{generalmethod}.
To find the asymptotics of 2D QWs on percolation graphs we expand an attractor space matrix in the natural basis
\begin{equation}
X_{\lambda, i} = \sum_{s_1,t_1,c_1,s_2,t_2,c_2} W^{s_1,t_1,c_1}_{s_2,t_2,c_2} | s_1, t_1, c_1 \rangle \langle s_2, t_2, c_2 |\,,
\label{2d:matrixelements}
\end{equation}
where $W^{s_1,t_1,c_1}_{s_2,t_2,c_2}$ denotes the matrix elements.
We also introduce the following shorthands corresponding to coin states $| L \rangle_C$, $| D \rangle_C$, $| U \rangle_C$, $| R \rangle_C$ acting
on indices corresponding to position states as
\begin{eqnarray}
 L(s,t) = (s-1,t)\,, & \quad D(s,t) = (s,t-1)\,,  \nonumber\\
 U(s,t) = (s,t+1)\,, & \quad R(s,t) = (s+1,t)\,,
\end{eqnarray}
for all $s,t$ which are well defined with respect to the given topology of the underlying graph, i. e. taking boundary conditions into account.
We define an involution $\sim$ acting on coin indices as $| \tilde{L} \rangle_C = | R \rangle_C$ and $ | \tilde{D} \rangle_C = | U \rangle_C$.

Using the notation we have just introduced, the shift conditions of
 (\ref{shift_conditions}) take the following form
\begin{eqnarray}
\label{2d:diagonalelements}
W^{c(s,t),c}_{c(s,t),c} & = & W^{s,t,\tilde{c}}_{s,t,\tilde{c}} \\
 \label{2d:antidiagonalelements}
 W^{s,t,\tilde{c}}_{c(s,t),c} & = & W^{c(s,t),c}_{s,t,\tilde{c}}\,.
\end{eqnarray}
Note, that indices $s,t,c$ run on their corresponding abstract space, i. e. $s,t$ on the sites of the 2D graph, and $c$ on coin state labels $L,D,U,R$. When $c(s_1,t_1) = (s_2,t_2)$ or $\tilde{c} = d$ is not satisfied all elements satisfy:
\begin{equation}
W^{c(s_1,t_1),c}_{d(s_2,t_2),d}  =  W^{s_1,t_1,\tilde{c}}_{s_2,t_2,\tilde{d}} =
W^{s_1,t_1,\tilde{c}}_{d(s_2,t_2),d}  =  W^{c(s_1,t_1),c}_{s_2,t_2,\tilde{d}}\,.
\label{2d:allelements}
\end{equation}
If $c(s_1,t_1)$ or $d(s_2,t_2)$ is not well defined, i. e. at least one of the points belong to a reflecting boundary (in the case of the carpet), the corresponding equations must be omitted from the set of equations defined above.
In summary, all matrix elements of an attractor space matrix satisfy (\ref{2d:diagonalelements}) and (\ref{2d:antidiagonalelements}), and non-diagonal matrix elements must satisfy the stricter condition of (\ref{2d:allelements}).
A natural way to solve these equations is first to find a subspace
satisfying (\ref{2d:allelements}) and then enlarge this subspace
allowing (\ref{2d:diagonalelements}) and
(\ref{2d:antidiagonalelements}).
Next, the condition imposed by 
(\ref{coin_condition}) must be solved on the now obtained abstract
subspace, resulting in the full attractor space.

In the following we show that with the help of pure
state ansatz proposed in Section \ref{purestatemethod}., the above strategy
can be simplified considerably.
The condition defining p-attractors (\ref{pureattractor_conditions}) for percolated QWs can be separated similarly as in the case of general attractors (\ref{separated_conditions}):
\begin{equation}
(I_P \otimes C) Y_{\lambda, i} (I_P \otimes C^{\dagger}) = \lambda S_{\mathcal{K}}^{\dagger} Y_{\lambda, i} S_{\mathcal{K'}}
\label{separated_pure_conditions}
\end{equation}
for all $\mathcal{K, K'} \subseteq E$.
Note, that in comparison with  (\ref{separated_conditions}), in  (\ref{separated_pure_conditions}) only the right side differs.
This implies, that the only difference comes from the shift conditions, namely:
\begin{equation}
S_{\mathcal{K}}^{\dagger}  Y_{\lambda, i} S_{\mathcal{K'}} = S_{\mathcal{L}}^{\dagger} Y_{\lambda, i} S_{\mathcal{L'}}
\end{equation}
for all $\mathcal{K, K', L, L'} \subseteq E$.
Considering this, the rule for matrix elements $V^{s_1,t_1,c_1}_{s_2,t_2,c_2} = \langle s_1, t_1, c_1  | Y_{\lambda, i} | s_2, t_2, c_2 \rangle $ of a possible p-attractor $Y_{\lambda, i}$
of 2D QWs can be obtained:
\begin{equation}
V^{c(s_1,t_1),c}_{d(s_2,t_2),d}  =  V^{s_1,t_1,\tilde{c}}_{s_2,t_2,\tilde{d}} =
V^{s_1,t_1,\tilde{c}}_{d(s_2,t_2),d}  =  V^{c(s_1,t_1),c}_{s_2,t_2,\tilde{d}}\,.
\label{2d:allelements_pure}
\end{equation}
which is the same rule as in  (\ref{2d:allelements}), but here it applies to all elements including diagonal elements.
Thus, our strategy is simplified further: First, we find all p-attractors. Second, we enlarge the found subspace by allowing
(\ref{2d:diagonalelements}) and (\ref{2d:antidiagonalelements}). In this way at least one additional attractor --- the trivial one, proportional to identity --- can be found.

Since p-attractors can be constructed from the common eigenstates of  a QW, they can be found rather easily.
These states are determined by the equations
\begin{equation}
I_P \otimes C | \phi_{\alpha, i_{\alpha}} \rangle = \alpha S_{\mathcal{K}}^{\dagger} | \phi_{\alpha, i_{\alpha}} \rangle
\label{common_eigenstates}
\end{equation}
for all $\mathcal{K} \subseteq E$.
First, like in the previous, the coin can be separated. Thus, all common eigenstates are confined to a subspace satisfying the shift conditions
\begin{equation}
 S_{\mathcal{K}}^{\dagger} | \phi_{\alpha, i_{\alpha}} \rangle =  S_{\mathcal{K'}}^{\dagger} | \phi_{\alpha, i_{\alpha}} \rangle
\end{equation}
for all $\mathcal{K, K'} \subseteq E$.
These shift conditions can be rewritten in the natural basis $| \phi \rangle = \sum_{s,t,c} \phi_{s,t,c} | s, t, c \rangle$, taking the elegant form of
\begin{equation}
 \phi_{s,t,\tilde{c}} = \phi_{\mathbf{c}(s,t),c} \,,
 \label{shiftconditions_vec}
\end{equation}
where the topology of the underlying graph (boundary conditions) again must be taken into account.
Second, in the subspace spanned by the latter shift conditions an ordinary and local eigenvalue problem determines the form of the coin states, and the possible spectrum:
\begin{equation}
I_P \otimes RC | \phi_{\alpha, i_{\alpha}} \rangle = \alpha  | \phi_{\alpha, i_{\alpha}} \rangle\,.
\label{spectrum}
\end{equation}
In the following we employ the just described procedure to explicitly solve certain percolated 2D  quantum walks.

\subsection{The 2D Hadamard walk: breaking of the directional symmetry}

The 2D Hadamard walk is a direct generalization of the 1D Hadamard walk, using the tensor product form coin
\begin{equation}
 H^{(2D)} = H \otimes H = \frac{1}{2} \left(
 \begin{array}{rrrr}
 1 & 1 & 1 & 1 \\
 1 & -1 & 1 & -1 \\
 1 & 1 & -1 & -1 \\
 1 & -1 & -1 & 1 \\
 \end{array}
 \right)\,,
 \label{Hadamardcoin}
\end{equation}
where
\begin{equation}
H  = \frac{1}{\sqrt{2}} \left(
 \begin{array}{rrrr}
 1 & 1 \\
 1 & -1  \\
 \end{array}
 \right)\,,
\end{equation}
is the well known coin operator of the 1D Hadamard walk.
In the undisturbed case this coin exhibits a spreading behaviour, which is characterized by peaks propagating from the origin at a constant velocity.  In the percolation case first we solve  $(\ref{spectrum})$ to gain the spectrum of p-states resulting in the set of eigenvalues $\{ \rmi, -\rmi, 1, 1 \}$. The corresponding eigenvectors of the $RC$ operator are $ | v_1 \rangle_C = \frac{1}{2} (1, -\rmi, -\rmi, -1)^T$, $ | v_2 \rangle_C = \frac{1}{2} (1, \rmi, \rmi, -1)^T$, $ | v_3 \rangle_C = \frac{1}{\sqrt{2}} (1, 0, 0, 1)^T$ and $ | v_4 \rangle_C = \frac{1}{\sqrt{2}}(0, 1, -1, 0)^T$, respectively. We find the following orthonormal basis on the percolated $M \times N$ carpet
\begin{eqnarray}
 \label{vector1}
 | \phi_1 \rangle & = & \sum_{s=0}^{M-1} \sum_{t=0}^{N-1} \frac{(-1)^s}{\sqrt{MN}} | s, t \rangle_P \otimes | v_1 \rangle_C\,, \\
  \label{vector2}
  | \phi_2 \rangle & = & \sum_{s=0}^{M-1} \sum_{t=0}^{N-1} \frac{(-1)^s}{\sqrt{MN}} | s, t \rangle_P \otimes | v_2 \rangle_C\,, \\
    \label{vector3}
 | \phi_3 (t) \rangle & = & \sum_{s=0}^{M-1}\frac{1}{\sqrt{M}} | s, t \rangle_P \otimes | v_3 \rangle_C\,, \\
  \label{vector4}
 | \phi_4 (s) \rangle & = & \sum_{t=0}^{N-1} \frac{(-1)^t}{\sqrt{N}} | s, t \rangle_P \otimes | v_4 \rangle_C\,.
\end{eqnarray}
The next step is to prove the completeness.
For the $\alpha = \rmi$ eigenvalue a general common eigenstate must have the form $| \phi \rangle = \sum_{s,t} a_{s,t} |s, t\rangle_P \otimes | v_1 \rangle_C$. Employing shift conditions  (\ref{shiftconditions_vec}) we get $a_{s+1,t} = - a_{s,t}$ and $a_{s,t+1} = a_{s,t}$, thus a single eigenvector is found and it takes the form (\ref{vector1}). Similarly, for $\alpha= -\rmi$ a single vector (\ref{vector2}) is found. For $\alpha = 1$ the general form of a common eigenstate is $ | \phi \rangle = \sum_{s,t} | s, t \rangle_P \otimes \left( a_{s,t} | v_3 \rangle_C + b_{s,t} | v_4 \rangle_C \right) $. Applying the shift conditions  we find $a_{s,t} = a_{s-1,t}$ and $b_{s,t} = -b_{s,t-1}$. This means $M+N$ free parameters, thus an $M+N$ dimensional subspace of common eigenstates with basis vectors (\ref{vector3}) and (\ref{vector4}).
Now, we have to determine the remaining attractors which cannot be constructed from common eigenstates.  We repeat, that from (\ref{2d:diagonalelements}) and (\ref{2d:antidiagonalelements}) we know that non p-attractors can be searched in a diagonal form.
Thus, solving the local equation (\ref{coin_condition}) for a diagonal coin block for $\lambda=1$ imposes
\begin{equation}
 \mathbf{B} = \left(
  \begin{array}{rrrr}
  a & c & C & A\\
  d & b & B & D\\
  -D & B & b & -d\\
  A & -C & -c & a
  \end{array}
 \right)\,,
\end{equation}
where $a - A = b+B$ and $D-d = c+ C$. As we require only diagonal coin blocks to be nonzero, and also for them $A=B=C=D=c=d=0$, thus $a=b$. This means that all diagonal coin blocks are proportional to identity $X_{s,t}^{s,t} = a_{s,t} I$. Due to shift conditions of  (\ref{2d:diagonalelements}) all $a_{s,t}$ are equal, thus a single attractor is revealed to be proportional to the identity, i. e. we found the trivial attractor as the only additional non-p attractor. Similarly, one can show that for the other possible attractor eigenvalues there are no additional non p-attractors. In summary, all attractors can be constructed employing (\ref{pureattractors}) and adding the trivial attractor proportional to identity.
Thus, the solution presented here is complete.

Let us have a look on the influence of boundary conditions on the available eigenvectors. It should be noted, that $| \phi_1 \rangle$ and $| \phi_2 \rangle$ are not available for periodic boundary condition in the $s$-direction with odd $M$. In a similar way $| \phi_4 \rangle$ is not a common eigenstate for periodic boundary condition in the $t$-direction with odd $N$.
From the $\alpha = \{ \rmi, -\rmi, 1, 1 \}$ pure state eigenvalues the possible attractor space eigenvalues are $\lambda = \{1, -1, \rmi, -\rmi\}$.
For the $\lambda =1$ eigenvalue, for all boundary conditions
\begin{eqnarray}
X_0 & = & I \\
X_1 (t_1,t_2) & = & | \phi_3 (t_1) \rangle \langle \phi_3 (t_2) |
\end{eqnarray}
are valid attractors, spanning a $1 + N^2$ dimensional space. For even $M$-s on periodic boundary conditions in the $s$ direction or open boundaries in the $s$ direction additional attractors
\begin{eqnarray}
X_2 & = & | \phi_1 \rangle\langle \phi_1 | \\
X_3 & = & | \phi_2 \rangle\langle \phi_2 |
\end{eqnarray}
form a two-dimensional space.
When in the $t$ direction the system is open or periodic with even $N$-s the additional
\begin{eqnarray}
X_4 (s_1, s_2) & = & | \phi_4  (s_1) \rangle\langle \phi_4 (s_2) | \\
X_5 (s_1,t_2) & = & | \phi_4 (s_1) \rangle\langle \phi_3 (t_2) | \\
X_6 (t_1,s_2)& = & | \phi_3 (t_1) \rangle\langle \phi_4 (s_2) |
\end{eqnarray}
attractors become available, forming an $M^2 + 2 MN $ dimensional space.

For the  superoperator eigenvalue $\lambda = \rmi$, for even $M$-s in the $s$
direction or open boundaries in the $s$ direction
\begin{eqnarray}
Y_1 (t_2) & = & | \phi_1 \rangle\langle \phi_3 (t_2) | \\
Y_2 (t_1) & = & | \phi_3 (t_1) \rangle\langle \phi_2 |
\end{eqnarray}
attractors are available spanning a $2N$ dimensional space.
The following attractors appear in addition if either we have open boundary condition in the direction $t$ or we have periodic boundary condition for $t$ with even $N$:
\begin{eqnarray}
Y_3 (s_2) & = & | \phi_1 \rangle\langle \phi_4 (s_2) | \\
Y_4 (s_1) & = & | \phi_4 (s_1) \rangle\langle \phi_2 |\,.
\end{eqnarray}
In that case the dimension of the attractor space is increased by $2M$.
The form of definition (\ref{attractor_space_matrices}) imply,
that the attractors corresponding to the conjugate $\lambda = -\rmi$
eigenvalue are simply the hermitian conjugate of the attractor space
matrices corresponding to $\lambda = \rmi$.

The last possible superoperator eigenvalue is $\lambda=-1$, with the attractors
\begin{eqnarray}
Z_1  & = & | \phi_1 \rangle\langle \phi_2 | \\
Z_2  & = & | \phi_2 \rangle\langle \phi_1 |
\end{eqnarray}
available when direction $s$ is open or periodic with even $M$,
adding a two-dimensional space to the attractor space. Altogether, the maximal number of
attractors for carpet (open boundaries) or for $even \times even$
torus (periodic boundaries) are $(M + N +2)^2 + 1$.

\begin{figure}[tb!]
\begin{center}
\begin{tabular}{cc}
\includegraphics[width=0.45\textwidth]{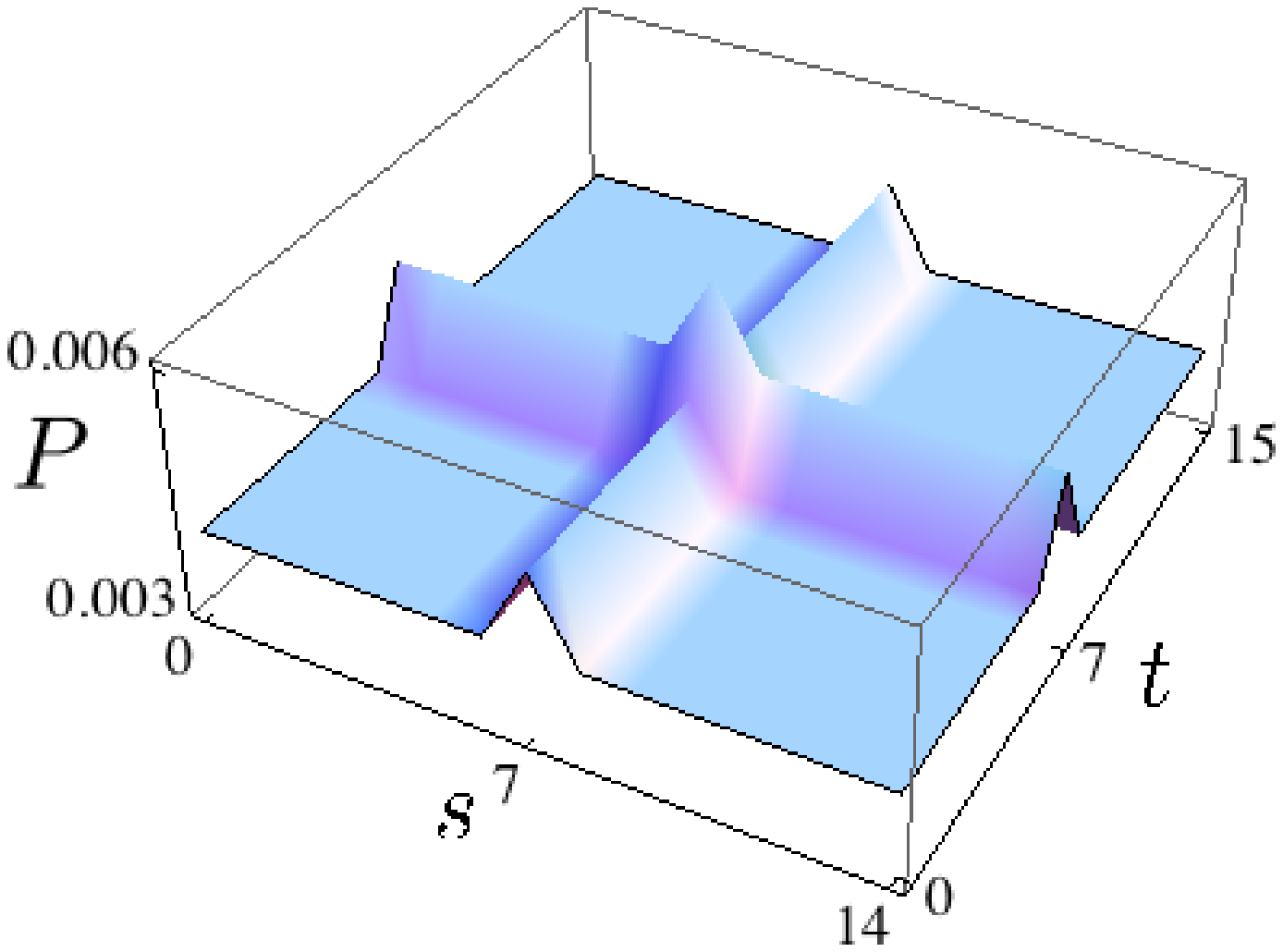} &
 \includegraphics[width=0.45\textwidth]{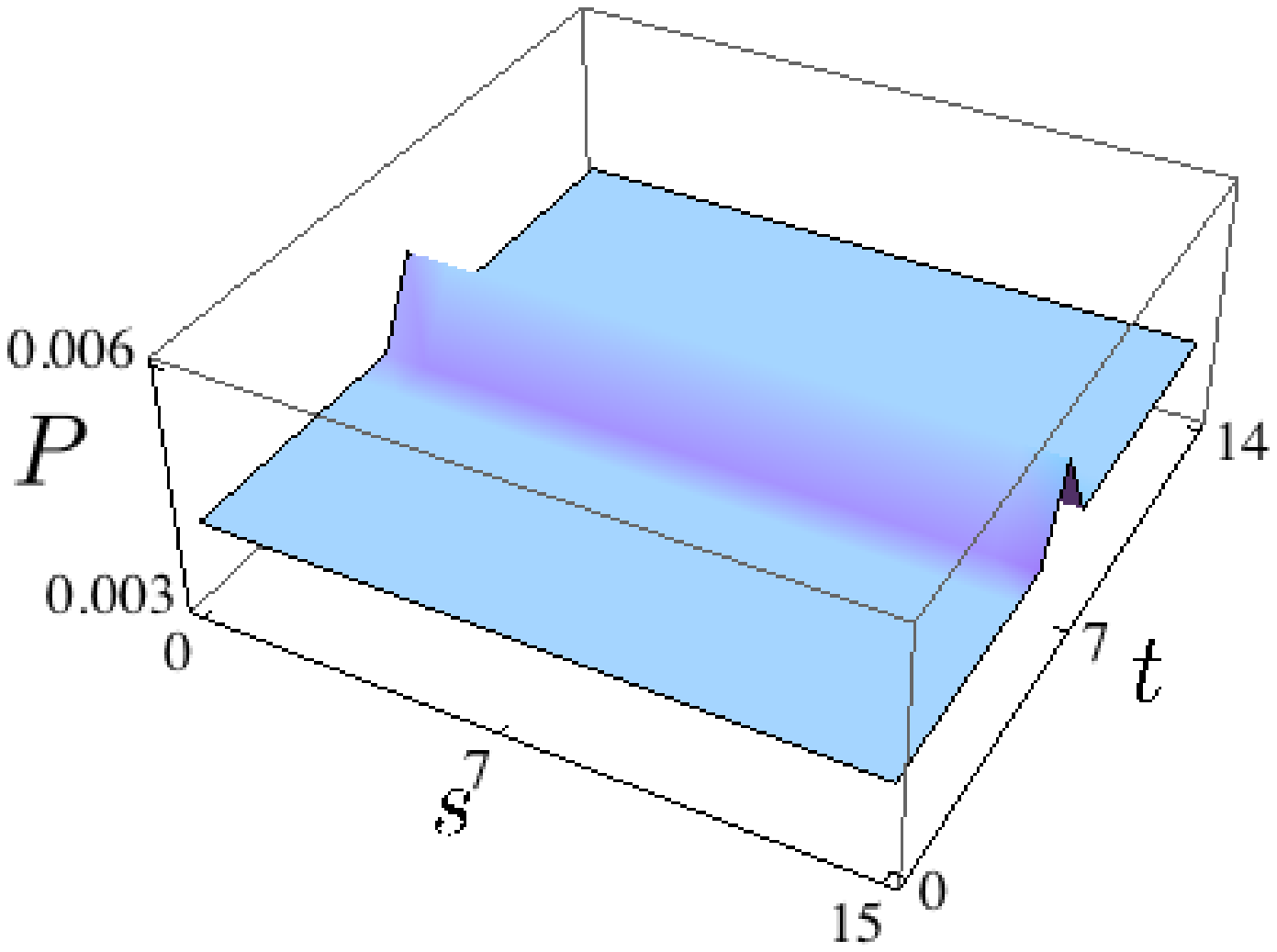}
\end{tabular}
\end{center}
\caption{ Asymptotic position probability distributions $P$ of
the 2D Hadamard walk on the torus graphs, starting from the
initial state: $ | 7,7 \rangle_P \otimes \frac{1}{\sqrt{2}}( |L
\rangle_C + | D \rangle_C )$. 
The left plot corresponds to the $15 \times 16$ percolation torus and carpet and the right plot corresponds to the $16 \times 15$ percolation torus. 
Due to the rotation of the underlying graph (and the initial state), the position distribution changes considerably.
However, in the case of carpets the position distribution rotates and remain otherwise unchanged.
The position distribution of the $15 \times 16$ percolation torus and carpet (rotated carpet) are identical (rotated). For the unpercolated (unitary) case the symmetry breaking is not observable within numerical precision.
 }
\label{hadamard_figure}
\end{figure}

Let us now
analyze the consequences one can draw from the explicit form of the eigenvectors (\ref{vector1}) - (\ref{vector4}) for the asymptotic behaviour
of the walks.
The common eigenvectors $| \phi_1 \rangle$ and $| \phi_2 \rangle$ in (\ref{vector1}), (\ref{vector2}) are uniform in position.
When the asymptotic state can be expanded by these, then the asymptotics will be uniform in position.
In contrast,
the other two families of eigenvectors $| \phi_3 (t) \rangle$ and $| \phi_4 (s)
\rangle$ in  (\ref{vector3}), (\ref{vector4}) are spatially non-uniform.
The asymptotic states built by them will have ridge like stripes.
Therefore, the boundary conditions for which $ | \phi_3 (t) \rangle$ or  $
| \phi_4 (s) \rangle$ are allowed can lead to  a non-uniform asymptotic position distribution.
While dynamical percolation means spatially a homogeneous source of decoherence, it may result
in a spatially inhomogeneous asymptotic distribution.

Further analyzing the asymptotically inhomogeneous solutions
we find, that percolation can cause the breaking of the directional symmetry, in the following sense.
Taking a certain initial state the unpercolated 2D Hadamard walk may show a directional symmetry for the position distribution:
if both the graph and the initial state are rotated by 90 degrees the resulting position distribution will also
be a rotated version of the original position distribution at all times.
In a numerical example we demonstrate that introducing percolation in this system can break the above directional symmetry.

Let us consider the example of a torus with size even $\times$ odd. A quantum walk with percolation
on such a torus will have an attractor space with dimension $(N+2)^2 + 1$.
In contrast, if we rotate the graph (odd $\times$ even torus) while keeping the coin operator the same,
we find an attractor space with dimension $(N+M)^2 +1$.
This change in the dimension of the attractor space clearly demonstrates the symmetry breaking.
Furthermore, by examining the corresponding eigenvalues we find that in the second case
(odd $\times$ even torus) only the $\lambda = 1$ eigenvalue occurs, leading to stationary asymptotic states.
Whereas in the first case (even $\times$ odd torus) also the $\lambda = \{-1, \pm \rmi\}$ eigenvalues will be included in the solution
possibly allowing for oscillations in the asymptotic state of the system.
In figure \ref{hadamard_figure}. we plot the asymptotic marginal position distributions, for the two cases.
Numerical simulations of a Hadamard walk on tori without percolation
show no difference between even $\times$ odd
and rotated odd $\times$ even systems within numerical precision. Thus, we conclude that the
directional symmetry breaking is induced by percolation. This
new phenomenon is one of the main results of the present paper.

\subsection{The Grover walk: trapping at the origin}
In this part we analyze the properties of the percolated walk
governed by the Grover diffusion operator $G_{i,j} = 2/d -
\delta_{i,j}$, that is
\begin{equation}
G = \frac{1}{2}  \left(
\begin{array}{rrrr}
-1 & 1 & 1 & 1 \\
1 & -1 & 1 & 1 \\
1 & 1 & -1 & 1 \\
1 & 1 & 1 & -1
\end{array}
\right)
\label{Grovercoin}
\end{equation}
for 2D QWs. The undisturbed Grover walk is characterized by the effect of trapping (localization), i. e. the probability of finding a particle in its initial position is not decaying to zero during the time evolution. In the 2D Grover walk there exists a single well defined localized initial coin state $| \psi^{\mathrm{spread}} \rangle = \frac{1}{2}(1,-1,-1,1)^T$ for which trapping effect is avoided, i. e. the wave function spreads freely.

In the following we determine the attractor space of the Grover
walk. We omit the details of the analytical proof as it is quite analogous to the
case of the Hadamard walk and hence can be easily reconstructed.
Similarly to the previous Hadamard case, we solve 
$(\ref{spectrum})$ to gain the spectrum of the common eigenstates.
After a lengthy, but straightforward calculation
we find the explicit form of the common eigenstates
\begin{eqnarray}
\fl | \phi_1 \rangle & = &  \frac{1}{\sqrt{4MN}} \sum_{s=0}^{M-1} \sum_{t=0}^{N-1} | s, t \rangle_P \otimes | v_1 \rangle_C\,, \\
\fl  | \phi_2 (s,t) \rangle & = & \frac{1}{\sqrt{8}} \big\{ | s, t \rangle_P \otimes | v_2 \rangle_C +  | s, t \oplus 1 \rangle_P \otimes \left( | v_2 \rangle_C + | v_3 \rangle_C \right) \nonumber\\ &&
  + | s \oplus 1, t \rangle_P \otimes \left( | v_2 \rangle_C + | v_4 \rangle_C \right) \nonumber\\ &&  +  | s \oplus 1, t \oplus 1 \rangle_P \otimes \left(  | v_2 \rangle_C +  | v_3 \rangle_C  +   | v_4 \rangle_C   \right) \big\} \,, \label{grover_loc_eigenstates} \\
 \fl | \phi_3 (s) \rangle & = & \sum_{t=0}^{N-1} \frac{(-1)^t}{\sqrt{2N}} | s, t \rangle_P \otimes | v_3 \rangle_C\,, \label{grover_phi3} \\
\fl | \phi_4 (t) \rangle & = & \sum_{s=0}^{M-1} \frac{(-1)^s}{\sqrt{2N}} | s, t \rangle_P \otimes | v_4 \rangle_C\,, \label{grover_phi4}
\end{eqnarray}
where
$ | v_1 \rangle_C = (1,-1,-1,1)^T$,  $ | v_2 \rangle_C = (1, 1,0,0)^T$, $ | v_3 \rangle_C = (0,-1,1,0)^T$ and $ | v_4 \rangle_C = (-1,0,0,1)^T$. By $\oplus$ we denote an addition
which takes the boundary conditions  into account.
For open boundary conditions (carpet) the part leaning over the boundary of the graph should be omitted (its amplitude is zero and the corresponding superposition state is normalized accordingly), and for periodic boundary conditions  the addition $\oplus$ corresponds to modulo operations with respect to the graph size.
These common eigenstates correspond to the eigenvalues $\alpha = \{-1, 1,1,1\}$, respectively.

\begin{figure}[tb!]
\begin{center}
\begin{tabular}{cc}
\includegraphics[width=0.45\textwidth]{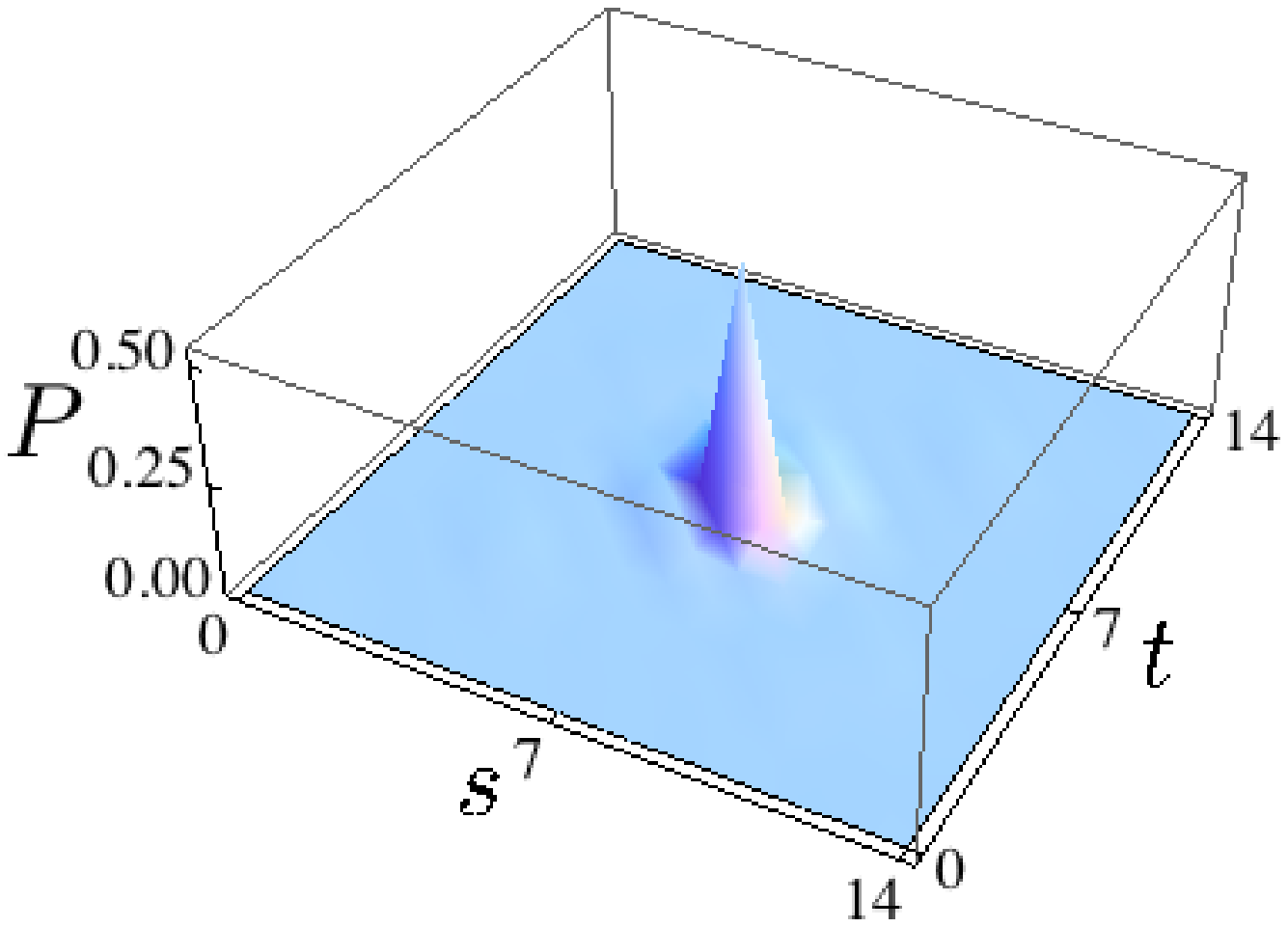} &
\includegraphics[width=0.45\textwidth]{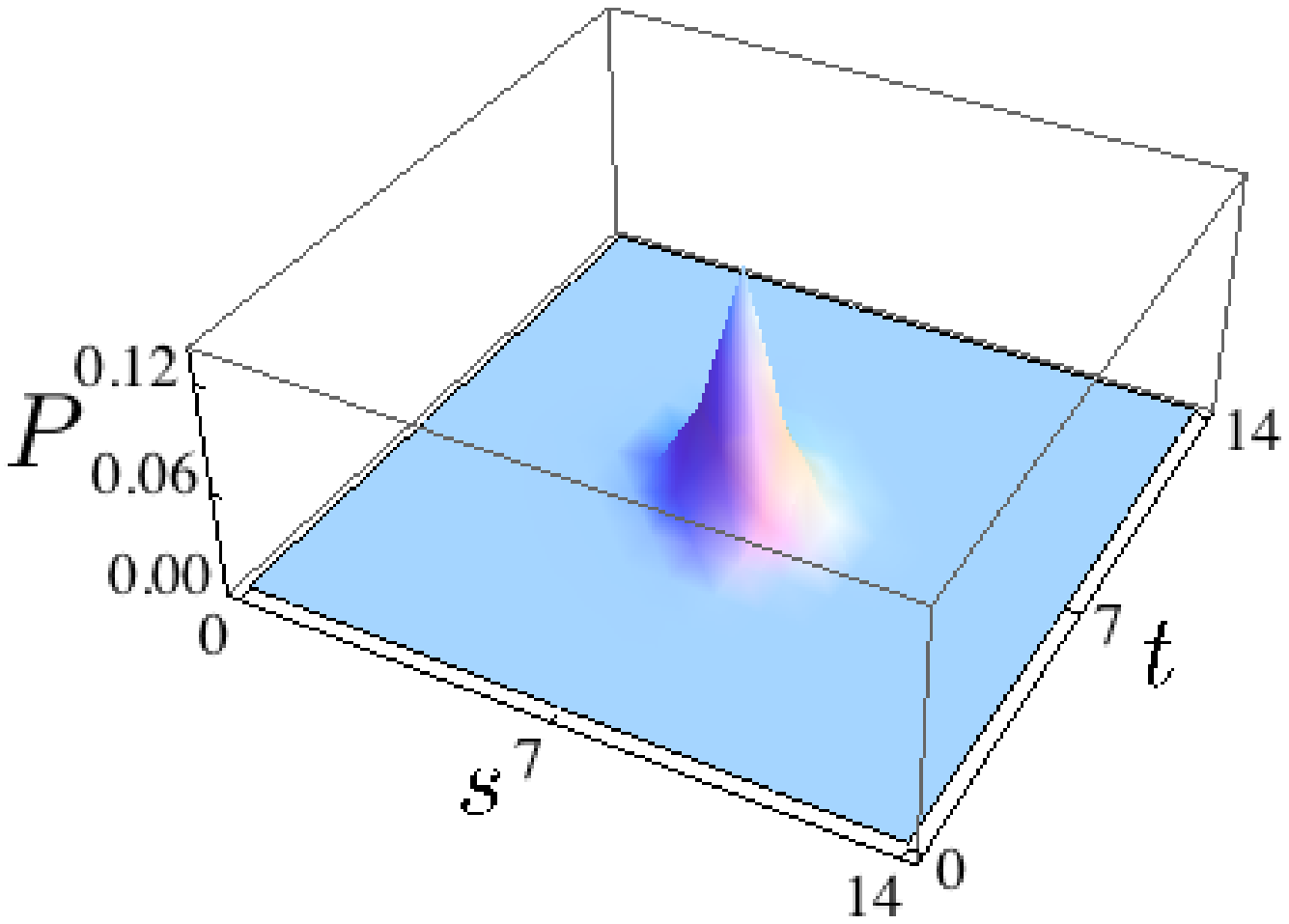}
\end{tabular}
\end{center}
\caption{ Position probability distribution $P$ of Grover
(\ref{Grovercoin}) walks on the $15 \times 15$ torus. The left plot
corresponds to 1000 steps of unitary time evolution (perfect graph)
started from $ | 7,7 \rangle_P \otimes \frac{1}{2}( |L \rangle_C + |
D \rangle_C + | U \rangle_C + | R \rangle_C )$ , whereas the right
plot corresponds to the asymptotic distribution on the percolation
case from the same initial state as the unperturbed one. It is
apparent that the trapping (localization) property of the walk is
preserved due to the common eigenstates
(\ref{grover_loc_eigenstates}) with finite support. We should point
out however that the localization effect is not as pronounced -- the
height of the peak is not as high as in the unperturbed case. }
\label{grover_figure}
\end{figure}

From the general theory we know, that all p-attractors can be
constructed form the above common eigenstates, employing 
(\ref{pureattractors}). More importantly, the attractors which do
not belong to the subspace of p-attractors should be found. However,
it turns out, that similarly to the Hadamard case the only non
p-attractor is the trivial one, proportional to identity.
Thus, the total number of attractors is $(MN+M+N+1)^2 +
1$ for all carpets, and $(MN+1)^2+1$ for tori where $M$ or $N$ are odd.
In the latter case (\ref{grover_phi3}) and (\ref{grover_phi4}) are restricted by the periodic boundary conditions --- they are not common eigenstates.
However, when $M$ and $N$ are both even in the case of tori, a single additional state from  (\ref{grover_phi3}) or (\ref{grover_phi4}) can be chosen as an additional common eigenstate, resulting in an attractor space with total number of attractors $(MN+2)^2+1$.
Note, that in all cases the attractor space is fully symmetric under rotations, thus in contrast to the 2D
Hadamard walk percolation induced symmetry breaking cannot be observed.

Analyzing the structure of the eigenstates we find another interesting effect related to trapping.
The common eigenstates $| \phi_2 (s,t) \rangle$  have finite support. The
importance of this observation is twofold: First, these states
cannot be sensitive to boundary conditions, thus one can expect that they remain common
eigenstates even on an infinite system. Second, these states
demonstrate the survival of the trapping (localization) effect: even
under the particularly strong decoherence effect of dynamical
percolation, the localized $| \phi_2 (s,t) \rangle$ states are invariants of the time evolution, thus they survive.
Consequently, an initially localized state overlapping with a $| \phi_2 (s,t)
\rangle$ state preserves its trapping property. A set of such robust
states may be perfect candidates for information storage, as they form
a decoherence free subspace and are separated spatially (except the neighbouring ones). This trapping effect for the percolation graph is illustrated in figure \ref{grover_figure}.

The above results motivate us to extend the definition of trapping.
The eigenstates with finite support, if they are independent of the boundaries represent trapping
in a general sense for quantum walks.  As we have seen, in particular cases
this general trapping property can survive decoherence effects in an open system.

\subsection{The Fourier walk}

In our last example we consider the Fourier coin operator as an example which has also been examined
in the context of 2D quantum walks. The Fourier coin does not have trapping initial states if started from a single position, but
its P\'olya number can be $1$, i. e. the walker can be recurrent  \cite{Stefanak2008a,Stefanak2008b}.

The Fourier coin in the natural basis takes the form of the discrete Fourier transformation matrix
\begin{equation}
F = \frac{1}{2}\left(
\begin{array}{rrrr}
1 & 1 & 1 & 1 \\
1 & -\rmi & -1 & \rmi \\
1 & -1 & 1 & -1 \\
1 & \rmi & -1 & -\rmi
\end{array}
\right)\,.
\label{DFTcoin}
\end{equation}
On the regular 2D lattice the undisturbed walk driven by the Fourier coin produces
a position probability distribution which is dominated by a slowly propagating peak and a ring like area. For a family of localized initial states $ | \psi^{\mathrm{ring}} \rangle_C = (a, b, a, -b)^T $ with $ |a|^2 + |b|^2 = \frac{1}{2} $ the central peak
disappears and one can observe a propagating ring like distribution.

Let us determine the attractor space of the Fourier walk on a 2D percolation lattice. The spectrum of common eigenstates is $\{ \alpha_n = \exp (\rmi \mu_n) \, | \, \mu_n = \frac{\pi}{8} (3 + 4n) \}$, and the common eigenstates themselves take the form of
\begin{equation}
\label{dft_purestates}
 | \phi_n \rangle = \sum_{s=0}^{M-1} \sum_{t=0}^{N-1} x^t_n y^s_n | s, t \rangle_P \otimes | v_n \rangle_C\,,
\end{equation}
where
\begin{equation}
\fl | v_n \rangle_C   =   (v_{n,1},v_{n,2},v_{n,3},v_{n,4})^T   =   \frac{1}{\mathcal{N}} \left( \left( \alpha_n + 1 \right)^2, \alpha_n^2 +\rmi, \alpha_n^2 - 1, 2 \alpha_n^3 + \alpha_n^2 -\rmi \right)^T
\end{equation}
with $\mathcal{N}^2 = 16 + 12 \cos(\mu_n) - 4 \sin (3 \mu_n)$, $x_n = v_{n,2} / v_{n,3}$ and $y_n = v_{n,1} / v_{n,4}$.
Since the derivation consists of very similar steps as we detailed for the 2D Hadamard case, we do not repeat them here.

From the structure of the common eigenstates it is apparent that these
states are not available on periodic boundary conditions ($M \times
N$ tori) except when both $M$ and $N$ are a multiple of $16$. The
p-attractors can be readily constructed by using 
(\ref{pureattractors}). The possible
asymptotic superoperator eigenvalues are $\lambda \in \{ 1, -1, \rmi, -\rmi \}
$. Now, we have to determine the number of non-p attractors.
Similarly to the previous cases, a straightforward analysis concludes
that the only non p-attractor available is the trivial one, which is proportional
to identity. Thus, the attractor space has the dimension of $16+1=17$ for
carpets, for $M \times N$ tori only the trivial identity is a valid
attractor, except when both $M$ and $N$ are a multiple of $16$.
In the latter case the attractor structure is concurrent with the carpet.

A careful second look at the states of 
(\ref{dft_purestates}) reveals that they
all correspond to a flat distribution in position. Thus, this coin induces a uniform asymptotic distribution and
the asymptotic limit cycles, determined by the superoperator
eigenvalues $\lambda \in \{-1, \rmi, -\rmi \} $ are only observable in
the coin degree of freedom. Nevertheless, the appearance of such
coins in the 2D cases by their connection with the similar behaviour observed for 1D walks \cite{Kollar2012} are
interesting.


All the previous case studies display quite different families of
asymptotic behaviours. The 2D Hadamard walk exhibits a percolation
induced symmetry breaking effect, which is an effect so far
unobserved for Hadamard walks. The system also shows ridge like
asymptotic states in position. This phenomenon is quite a departure from the
always uniform class of 1D QWs \cite{Kollar2012} including the Hadamard coin. The Grover walk keeps its trapping
(localization) property despite of the strong decoherence caused by
the dynamical percolation. The last showcased walk (Fourier coin) shows
nontrivial behaviour only in the internal degree of freedom.
The position distribution is uniform similarly to the above class of 1D walks.
It is important to note, that all these effects are controlled only
by the selection of the coin operator.

\section{Conclusions}
Quantum walks can be used to model a variety of processes. The ideal
walk is described by a unitary evolution. Disturbances of this ideal
description lead to decoherence with a number of unexpected new
effects. In the present paper we studied a special case of
decoherence caused by dynamically broken links of the underlying graph defining
the position space of the walk. This kind of perturbation is referred to as dynamical percolation.

Using the method of random unitary operations we solved the
asymptotic dynamics of percolated  two-dimensional walks. We
have extended the method sketched in our previous paper
\cite{Kollar2012} and have shown that explicit analytic forms can be obtained.
This is facilitated by the fact that a considerable part of the attractor space elements can be
determined using the pure eigenstate ansatz.
Here we emphasize that this ansatz  is a universal tool applicable for
general RUO maps to determine a decoherence free subspace, however, it does not
necessarily provide us with the complete solution for the attractors space.

Based on the presented analytic tools we derived a number of results
which are characteristic for percolated quantum walks in general and
discussed some of the results which are dependent on the topology (boundaries)
of the underlying graph. Using the general formalism we have proven
that for all percolated quantum walks the asymptotic position
distribution is always stationary in time. For
special choices of the boundary conditions we showed that
percolation induced symmetry breaking can be observed for a walk
driven by the 2D Hadamard coin. Our extensive study also revealed
position non-uniformity in the asymptotics, an effect which is not possible for 1D Hadamard walks on percolation graphs.
For the two-dimensional Grover walk we observed trapping caused by the existence
of a family of robust eigenstates with finite support, that form a
decoherence free subspace. Based on this observation one can extended
the concept of trapping (localization) to finite or open systems. All the
asymptotic properties of the studied systems can be controlled by
the careful selection of coin, thus the results of the present paper
motivate further studies on the classification of coin operators.

These newly discovered effects may be demonstrated using present
state experiments. The considerable degree of control on the walk is
within reach of optical feedback loop experiments implementing
quantum walks \cite{Schreiber2012}. They allow dynamic control
over the quantum walk topology which is the basic requirement for
the observation of the just described phenomena.

\ack

We acknowledge support by MSM 6840770039, GACR 13-33906S, RVO 68407700, the
Hungarian Scientific Research
Fund (OTKA) under Contract Nos. K83858, NN109651,
the Hungarian Academy of Sciences (Lend\"ulet Program,
LP2011-016).
B.~K. acknowledges support
by the European Union and the State of Hungary, co-financed by the European Social Fund in the framework of T\'AMOP 4.2.4. A/2-11-1-2012-0001 "National Excellence Program".

\end{document}